\documentclass[twocolumn,aps,superscriptaddress]{revtex4-1} 
\usepackage{amsmath,amssymb,graphicx,times,color,upgreek}
\usepackage[colorlinks, linkcolor=blue, citecolor=blue, urlcolor=blue, breaklinks=true]{hyperref}
\usepackage{amsmath}
\usepackage{lettrine}
\usepackage{amsmath}
\usepackage{graphicx}
\usepackage{dcolumn}
\usepackage{bm}
\usepackage{textcomp}
\usepackage{xcolor}

\newcommand{\beq}{\begin{equation}}
\newcommand{\eeq}{\end{equation}}

\begin{document}

\title{Microwave quantum illumination
using a digital %phase-conjugate 
receiver}

\author{S. Barzanjeh}
\email{shabir.barzanjeh@ist.ac.at}
\address{Institute of Science and Technology Austria, am Campus 1, 3400 Klosterneuburg, Austria}

\author{S. Pirandola}
\address{Department of Computer Science, University of York, Deramore Lane, York YO10 5GH, United Kingdom}
\address{Research Laboratory of Electronics, Massachusetts Institute of Technology, Cambridge MA 02139, USA}

\author{D. Vitali}
\address{School of Science and Technology, Physics Division, University of Camerino, Camerino (MC), Italy}
\address{INFN, Sezione di Perugia, Italy}
\address{CNR-INO, Firenze, Italy}

\author{J. M. Fink}
\email{jfink@ist.ac.at}
\address{Institute of Science and Technology Austria, am Campus 1, 3400 Klosterneuburg, Austria}

\date{\today}
\begin{abstract}
Quantum illumination is a powerful sensing technique that employs entangled signal-idler photon pairs to boost the detection efficiency of low-reflectivity objects in environments with bright thermal noise. The promised advantage over classical strategies is particularly evident at low signal powers, a feature which could make the protocol an ideal prototype for non-invasive biomedical scanning or low-power short-range radar. In this work we experimentally investigate the concept of quantum illumination at microwave frequencies. We generate entangled fields using a Josephson parametric converter to illuminate a room-temperature object at a distance of 1 meter in a free-space detection setup. We implement a digital phase conjugate receiver based on linear quadrature measurements that outperforms 
a symmetric classical noise radar in the same conditions despite the entanglement-breaking signal path. Starting from experimental data, we also simulate the case of perfect idler photon number detection, which results in a quantum advantage compared to the relative classical benchmark. Our results highlight the opportunities and challenges on the way towards a first room-temperature application of microwave quantum circuits.
\end{abstract}
               
\maketitle

Quantum sensing is well developed for photonic applications~\cite{SensingReview} inline with other advanced areas of quantum information~\cite{NielsenChuang,Weedbrook12,Hayashi17,Watrous18}. As a matter of fact, quantum optics has been so far the most natural and convenient setting for implementing the majority of protocols in quantum communication, cryptography and metrology~\cite{Braunstein94}. The situation is different at longer wavelengths, such as THz or microwaves, for which the current variety of quantum technologies is more limited and confined to cryogenic environments. With the exception of superconducting quantum processing~\cite{Schoelkopf08} no microwave quanta are typically used for applications such as sensing and communication. For such tasks high energy and low loss optical and telecom frequency signals represent the first choice and form the communication backbone in the future vision of a hybrid quantum internet~\cite{Kimble2008,PirBra16,Wehner18}.

Despite this general picture, there are applications of quantum sensing that are naturally embedded in the microwave regime. This is exactly the case with quantum illumination (QI)~\cite{Lloyd08,Tan08,Lopaeva2013,Zhang2013,Zhang2015,Weedbrook2016,Shapiro2019} for its remarkable robustness to background noise, which at room temperature amounts to $\sim 10^3$ thermal quanta per mode
at a few GHz. In QI, the aim is to detect a low-reflectivity object in the presence of very bright thermal noise. This is accomplished by probing the target with less than one entangled photon per mode, in a stealthy non-invasive fashion which is impossible to reproduce with classical means. In the Gaussian QI protocol~\cite{Tan08}, the light is prepared in a two mode squeezed vacuum state~\cite{Weedbrook12} with the signal mode sent to probe the target while the idler mode is kept at the receiver. Even though entanglement is lost in the round trip from the target, the surviving signal-idler correlations, when appropriately measured, can be strong enough to beat the performance achievable by the most powerful classical detection strategy. In the low photon flux regime, where QI shows the biggest advantage, it could be suitable for extending quantum sensing techniques to short-range radar~\cite{Choi2016} and non-invasive diagnostic scanner applications~\cite{Lin1992}.

Previous experiments
%relied directly on the measured quadratures and 
in the microwave domain~\cite{Wilson19, Wilson1903} demonstrated a quantum enhancement of the detected covariances
%enabled improvement of the signal-to-noise ratio 
compared to a symmetric classical noise radar, i.e.~with approximately equal signal and idler photon number. With appropriate phase sensitive detection an ideal classically correlated noise radar can be on par or, in the case of a bright idler \cite{Shapiro2019}, even outperform coherent heterodyne detection schemes, which maximize the signal-to-noise (SNR) ratio for realistic (phase-rotating) targets.
%
%In general, an ideal classical noise radar can potentially be on par with 
%%or, in the case of a very strong signal - idler imbalance and appropriate detection, 
%or might even be able to outperform coherent heterodyne detection schemes, which
%%with heterodyne detection of a coherent signal, the most powerful classical strategy in terms of 
%maximize the signal to noise ratio for realistic (generally phase-rotating) targets. 
%
However, if the phase of the reflected signal is stable over relevant timescales or a priori known, homodyne detection represents the 
%to gain an additional 3?dB in the SNR which represents 
strongest classical benchmark. 
%In the low photon flux regime, where QI shows the biggest advantage, it could be suitable for extending quantum sensing techniques to short-range radar and non-invasive diagnostic scanner applications, e.g. for healthcare monitoring \cite{Lin1992}, security, smart homes, ground-penetrating radar, and autonomous driving~\cite{Choi2016}.}

%%%%%%%%%%%%%%%

\begin{figure*}[t]
\centering
\includegraphics[width=1.7\columnwidth]{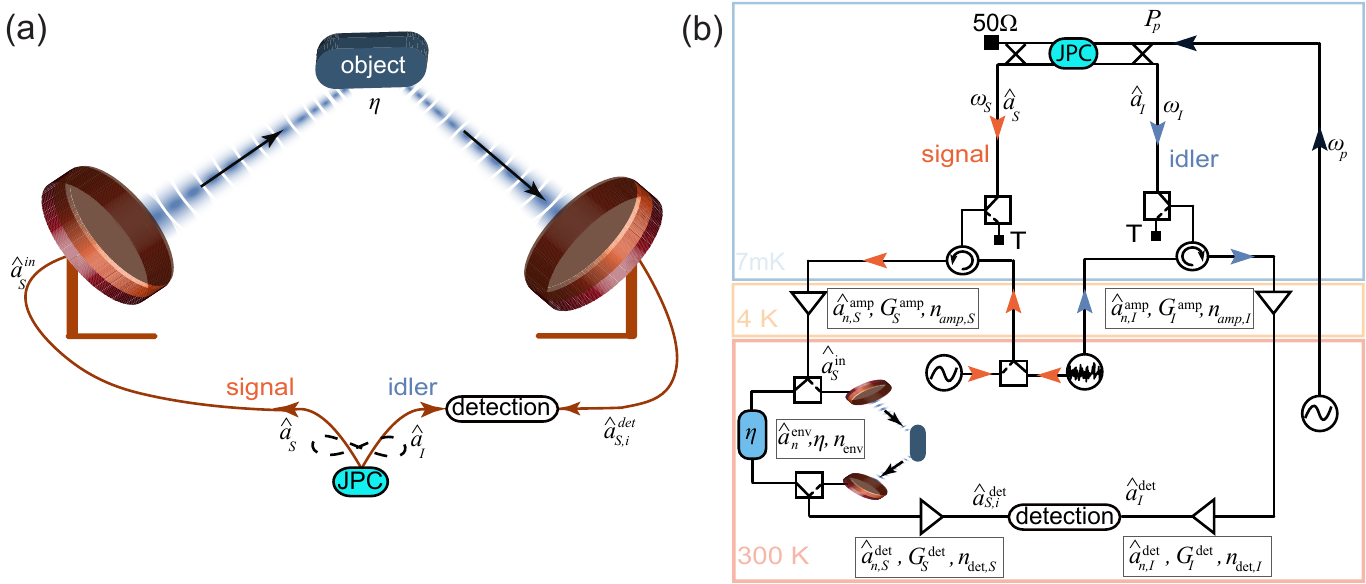}
\caption{\textbf{Implementation of microwave quantum illumination.} (a) Schematic representation of microwave quantum illumination. A quantum source generates and emits stationary entangled microwave fields in two separate paths. The signal mode $\hat a_S$ is used to interrogate the presence ($i=1$) or absence ($i=0$) of a room-temperature object with total round-trip reflectivity $\eta$. The returned mode $\hat a_{S,i}^\mathrm{det}$ is measured together with the unperturbed idler mode $\hat a_I$. (b) Circuit diagram of the experimental setup. A superconducting Josephson parametric converter (JPC) is used to entangle signal and idler modes at frequencies $\omega_S$ and $\omega_I$  by applying a suitable parametric pump tone at the sum frequency $\omega_p=\omega_S+\omega_I$ at $\sim$~7~mK. 
A coherent microwave tone or a classically correlated noise source are used to generate benchmark signals at room temperature that are sent into the dilution refrigerator and reflected from the JPC ports.
The outputs of the JPC or the reflected classical signals are amplified, down-converted and digitized simultaneously and independently for both channels. The signal mode passes through a measurement line that contains a room-temperature switch that is used to select between a digitally controllable attenuator $\eta$
%representing an object with tunable reflectivity $\eta$
and a free-space link realized with two antennas and a movable reflective object. Here, we consider $\eta$ as the total signal loss between the two room temperature switches used in our measurement chain. For the system noise and gain calibration, we use
%a pair of copper coaxial cables to
two latching microwave switches at cold temperatures which are
used to select between the JPC outputs and a temperature $T$ variable $50\,\Omega$ load (black squares).  %Classical illumination and coherent-state illumination are performed using noise sources and coherent states at room-temperature.
In both panels above, the final detection step corresponds to a 2 channel quadrature measurement followed by digital post-processing.} \label{Fig1}
\end{figure*}

In this work, we implement a digital version of the phase-conjugated receiver of Ref. \cite{Guha}, experimentally investigating proof of concept QI in the microwave regime~\cite{Barzanjeh15}. We use a Josephson parametric converter (JPC) \cite{Bergeal2010, Abdo2013} inside a dilution refrigerator for entanglement generation \cite{Flurin2012,Silveri2015}. The generated signal microwave mode, with annihilation operator $\hat a_S$, is amplified to facilitate its detection and sent to probe a room-temperature target, while the idler mode $\hat a_I$ is measured as schematically shown in Fig.~1(a). The reflection from the target $\hat a_R$ is also detected and the two measurement results are post-processed to calculate the SNR for discriminating the presence or absence of the object. Our experimental implementation of QI relies on linear quadrature measurements and suitable post-processing in order to compute all covariance matrix elements from the full measurement record as shown in previous microwave quantum optics experiments with linear detectors \cite{daSilva10,Bozyigit11,Eichler12}. This enables an implementation of the phase-conjugated receiver that fully exploits the correlations of the JPC output fields without analog photodetection. We then compare the SNR with other detection strategies for the same signal path, i.e. the same signal photon numbers at the JPC output, which is also our reference point for the theoretical modeling. 

Our digital approach to QI circumvents common practical problems such as finite idler storage time that can limit the range and fidelity of QI detection schemes. However, this advantage comes at the expense that the theoretically strongest classical benchmark in the same conditions - the coherent state homodyne detector using the same signal power and signal path - can be approached in specific conditions such as quantum limited amplification, but never be outperformed. To outperform coherent state homodyne detection in practice, will require low temperature square law detection of microwave fields that can be realized with radiometer or photon counting measurements. Nevertheless, 
% and allows to surpass the classical benchmark at low signal photon flux. 
%Using careful calibration measurements of the idler path allows to simulate a situation with perfect idler photon number detection, i.e. resembling the situation where the reflected mode is detected together with the idler mode using analog microwave photon counters. 
using calibration measurements of the idler path, we can simulate a situation with perfect idler photon number detection, extrapolating the case where the reflected mode is detected together with the idler mode using analog microwave photon counters.
For this situation we show that the SNR of coherent heterodyne detection and symmetric noise radars is exceeded by up to 4~dB and that of homodyne detection - the classical benchmark - by up to 1~dB for the same amplified signal path, measurement bandwidth and signal power. 
%To make use of this advantage in practice, will require either quantum limited heterodyne measurements of the idler signal, which would be sufficient to outperform the coherent heterodyne detection, or actual low temperature microwave photon counting measurements required to outperform the coherent homodyne detector.
We also note that the strong and noisy amplification of the signal path chosen to facilitate the detection with commercial analog-to-digital converters
%used for short range detection with low signal loss 
enables another classical receiver strategy, i.e.~the detection of the amplifier noise in the presence of the target. Since the amplified noise exceeds the environmental noise at room temperature by orders of magnitude, this would indeed be the most effective strategy for the implemented experiment. For the same reason, a low noise coherent source at room temperature would outperform the relative benchmarks considered here. In practice, outperforming the room temperature benchmark depends on the chosen amount of gain, the type of amplifier and the loss in the detection system and therefore does not pose a fundamental limitation to the presented measurement scheme that focusses on the relative comparison of the different illumination types.

% Entangled microwave signal and idler photons have already been used in ~\cite{Wilson19, Wilson1903}, where however the two beams have been measured independently, without fully exploiting therefore the existing quantum correlation. We show here quantum advantage in the detection of a copper plate target in a range of about 1 meter, demonstrating therefore a first step for the exploitation of quantum features in radar detection. Quantum illumination yields its best performance in the low signal-to-noise regime and therefore our results are promising for short range non-invasive applications, e.g. security checks, and target detection in biological tissues. For long-distance more standard radar applications quantum illumination could in principle also provide range and velocity information still using classical time-of-flight and Doppler techniques, but the prospects are much less evident.}
The experimental setup, shown in Fig.~1(b), is based on a frequency tunable superconducting JPC operated in the three-wave mixing regime and pumped at the sum of signal and idler frequencies $\omega_p=\omega_S+\omega_I$, see Methods for more details. The output of the JPC contains a nonzero phase-sensitive cross correlation $\langle \hat a_S \hat a_I\rangle$, which leads to entanglement between the signal mode with frequency $\omega_S= 10.09$~GHz and the idler mode with frequency $\omega_I= 6.8$~GHz. In our work, the quantities $\langle \hat O\rangle$ and  $(\Delta O_i)^2=\langle\hat{O}_i^2\rangle-\langle\hat{O}_i\rangle^2$ define the mean and the variance of the operator $\hat{O}$, respectively, and they are evaluated from experimental data. The signal and idler are sent through two different measurement lines, where they are amplified, filtered, down-converted to an intermediate frequency of $20$ MHz and digitized with a sampling rate of 100 MHz using an 8 bit analog-to-digital converter. Applying fast Fourier transform and post-processing to the measured data, we obtain the quadrature voltages $I_i$ and $Q_i$, which are related to the complex amplitudes $a_i$ and their conjugate $a_i^*$ of the signal and idler modes at the outputs of the JPC as $a_i=\frac{I_i+\mathrm{i}\, Q_i}{\sqrt{2\hbar \omega_i B R G_i}}$ and $a_i^*=\frac{I_i-\mathrm i\, Q_i}{\sqrt{2\hbar \omega_i B R G_i}}$, having the same measurement statistics as the annihilation operator $\hat{a}_i$. Here, $R=50\,\Omega$, $B=200$ kHz is the measurement bandwidth set by a digital filter and $i=S,I$ \cite{Menzel,Eichler12,Barzanjeh19}. We calibrate the system gain $(G_S,G_I)=(93.98(01),94.25(02))$ dB and system noise $(n_{\mathrm{add},S},n_{\mathrm{add},I})=(9.61(04),14.91(1))$ of both measurement channels as described in Methods.

A first important check for the experiment is to quantify the amount of entanglement at the output of the JPC at 7
mK. A sufficient condition for the signal and idler modes to be entangled is the non-separability criterion~\cite{Duan2000} $\Delta:=\langle \hat X^2_{-}\rangle+\langle \hat
P^2_{+}\rangle<1$, for the joint field
quadratures $\hat X_{-}=(\hat a_S+\hat a_S^\dagger-\hat a_I-\hat
a_I^\dagger)/2$ and  $\hat P_{+}=(\hat a_S-\hat a_S^\dagger+\hat
a_I-\hat a_I^\dagger)/(2\mathrm{i})$. % possess strong quantum correlations.
In Fig.~2(a) we show measurements of
$\Delta$ as a function of the signal photon number
$N_S=\langle \hat a_S^\dagger \hat a_S\rangle$ at the output of
the JPC at millikelvin temperatures, as obtained by applying the above calibration procedure to both signal and idler modes,
%and the pump power $P_p$ at the input of the JPC 
and compare the result with classically-correlated
radiation. The latter is generated at room temperature using the white
noise mode of an arbitrary waveform generator, divided into two different lines, individually
up-converted to the signal $\omega_S$ and idler
$\omega_I$ frequencies and fed to the JPC inside
the dilution refrigerator. Note that, for both JPC and classically correlated noise, we digitally rotate the relative phase of the quadratures to maximize the correlation between signal and idler.
%\textcolor{blue}{In both cases we use phase rotation of the data to minimize $\Delta_\mathrm{EPR}$.}
\begin{figure}[t]\label{Fig2}
\centering
\includegraphics[width=\columnwidth]{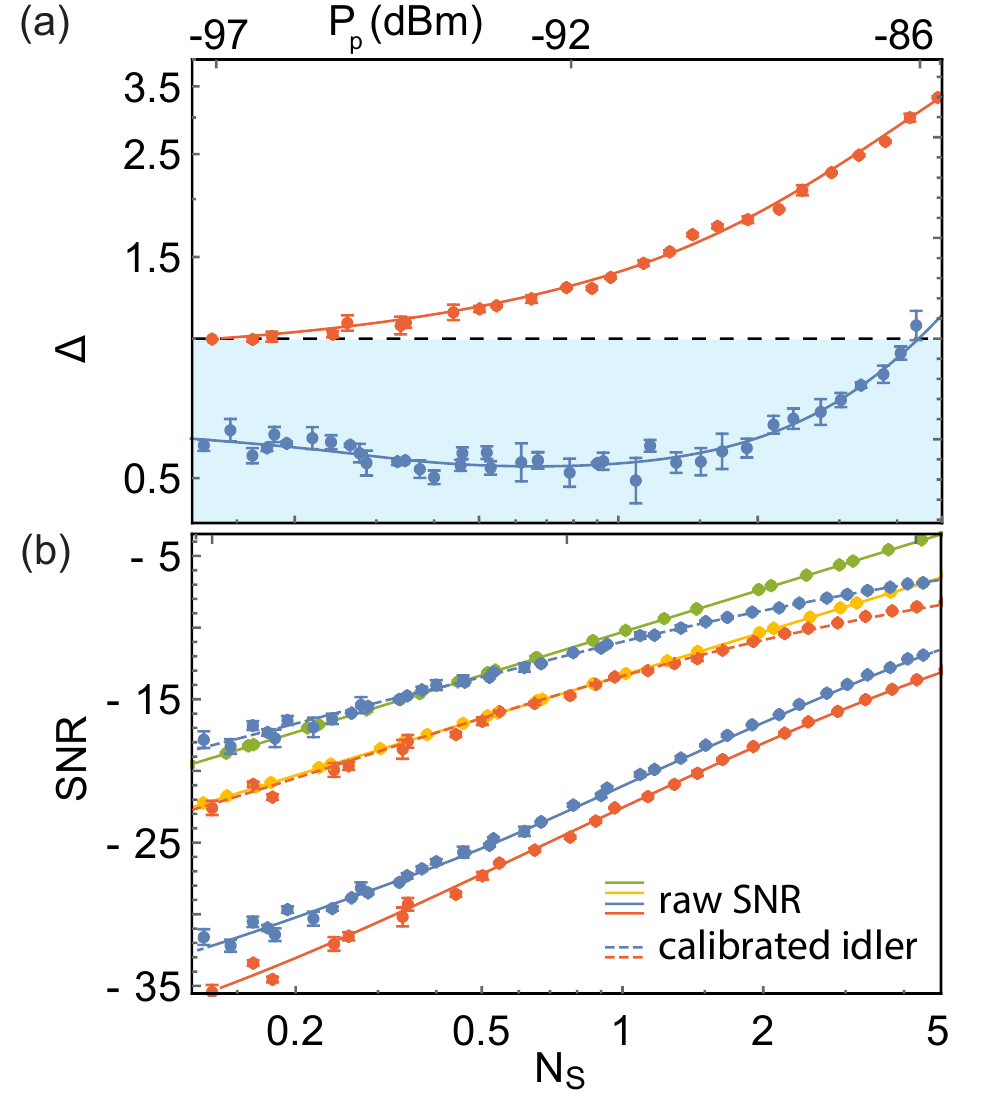}
\caption{\textbf{Entanglement and quantum illumination.} (a) The measured entanglement parameter $\Delta$ for the output of the JPC (blue) and classically-correlated noise (red) as a function of the inferred signal photon number $N_S$ at the output of the JPC and the pump power $P_p$ at the input of the JPC. 
(b) Comparison of the measured single mode signal-to-noise ratio (SNR) of quantum illumination (QI, solid blue), symmetric classically-correlated illumination (CI, solid red), coherent-state illumination with homodyne (solid green) and heterodyne detection (solid yellow), and the inferred SNR of calibrated QI (dashed blue) and CI (dashed red) as a function of the signal photon number $N_S$ for a perfectly reflective object and a $5\,\mu$s measurement time. The dots are measured and inferred data points and the solid and dashed lines are the theory prediction.
For both panels (a) and (b) the error bars indicate the $95\%$ confidence interval based on 3 sets of measurements, each with $380\,$k two channel quadrature pairs for QI/CI, and $192\,$k quadrature pairs for coherent-state illumination.}
\end{figure}

The classically-correlated signal and idler
modes are then reflected back from the JPC (pumps are off) and pass
through the measurement lines attached to the outputs of the JPC.
This ensures that both classical and quantum radiations
experience the same conditions in terms of gain, loss, and noise
before reaching the target and before being detected in the identical way. As shown in Fig.~2(a), at low photon
number the parameter $\Delta$ is below one proving
that the outputs of the JPC are entangled, while at larger photon
number (larger pump power) the entanglement gradually degrades and
vanishes at $N_S=4.5$ photons $\cdot \mathrm{s^{-1}\cdot Hz^{-1}}$. 
We attribute this to finite losses in the JPC, which leads to pump power dependent heating and results in larger variances of the output field. The classically correlated radiation of the same signal power on the other hand (red data points), cannot fulfill the non-separability criterion and therefore $\Delta\geq 1$ for the entire range of the signal photons. In the latter case we also observe a slow relative degradation of the classical correlations as a function of the signal photon numbers, which could be improved with more sophisticated noise generation schemes \cite{Wilson19}. 
%\textcolor{blue}{Unclear to me, in which sense we believe we can improve a classical benchmark. We know it is CS+hom and we know classical noise radar is not bad but it is not the benchmark}.

%In the low photon number regime the classical radiation is near the vacuum state
%$\Delta_\mathrm{EPR}\simeq 1$ while at larger photon number the
%state clearly becomes thermal  $\Delta_\mathrm{EPR}> 1$.

%We experimentally realize the described quantum illumination
%platform in the microwave domain using the JPC and compare its
%performance to the classically-correlated and coherent state
%illuminations.

The experiments of QI and classically-correlated illumination (CI)
%, i.e., based on a classically-correlated source, 
are implemented in a similar way, see Fig. \ref{Fig1}(b). The two amplified quadratures of the idler mode 
%$\hat a_I$ is measured and the outcome 
$\hat a_I^\mathrm{det}$ are measured at room temperature, and
%Using the system noise and gain, extracted from the calibration measurement, we accurately post-process $\hat a_I^\mathrm{det}$ used for the reconstruction of the covariance matrix of the detected signal $\hat a_S^\mathrm{det}$ and inferred idler state $\hat a_I$ directly at the output of the JPC at a temperature of $\sim$7~mK.} The noise calibration of the idler is vital to observe a quantum advantage, because the amplifier added noise directly contributes to the total noise of the entire detection process. 
the signal mode $\hat a_{S}$ 
%on the other hand 
is amplified (with gain $G_S^{\mathrm{amp}}$ and the noise mode $\hat a_{n,S}^{\mathrm{amp}}$) and used to probe a noisy region that is suspected to contain an object. In this process, we define $\eta$ as the total detection loss on the signal path between the two room-temperature switches used in the measurement chain, which includes cable loss, free-space loss, and object reflectivity.
The reflected signal from the region is measured, by means of a mixer and an amplifier with gain $G_S^{\mathrm{det}}$ and the noise mode $\hat a_{n,S}^{\mathrm{det}}$. The output  $\hat a_{S,i}^{\mathrm{det}}$ in the presence ($i=1$) or absence ($i=0$) of the object is then post-processed for the reconstruction of the covariance matrix of the detected signal-idler state. 
%mode (without noise calibration) and the inferred idler mode (with noise calibration)}.

The signal mode $\hat a_{S,i}^{\mathrm{det}}$ takes different forms depending on the presence
%is detected without applying any gain and noise calibration.
%We consider an equally likely binary condition in which the region either contains (hypothesis $H_1$, $i=1$) or does not contain (hypothesis $H_0$, $i=0$) a low-reflectivity object.
\begin{eqnarray}\label{eq5}
\hat a_{S,1}^{\mathrm{det}}=\sqrt{G_S}\Big(\sqrt{\eta}\hat a_S&&+\sqrt{\frac{\eta(G_S^{\mathrm{amp}}-1)}{G_S^{\mathrm{amp}}}}\hat a_{n,S}^{\mathrm{amp\dagger}}+\sqrt{\frac{1-\eta}{G_S^{\mathrm{amp}}}}\hat a_{n}^{\mathrm{env}}\nonumber\\
&&+\sqrt{\frac{G_S^{\mathrm{det}}-1}{G_S}}\hat a_{n,S}^{\mathrm{det\dagger}}\Big),
\end{eqnarray}
or absence
\begin{equation}\label{eq6}
\hat a_{S,0}^{\mathrm{det}}=\sqrt{G_{S}^{\mathrm{det}}}\Big(\hat a_{n}^{\mathrm{env}}+\sqrt{ 1-\frac{1}{G_S^{\mathrm{det}}}}\hat a_{n,S}^{\mathrm{det}\dagger}\Big),
\end{equation}
of the target with $\hat a_{n}^{\mathrm{env}}$ is the environmental noise mode. In the absence of the object, the signal contains only noise $n_{0}= G_{S}^{\mathrm{det}} n_\mathrm{env}+(G_{S}^{\mathrm{det}} -1)n_{\mathrm{det},S}$ in which $n_{\mathrm{det},S}$ is the amplifier added noise after interrogating the object region. In the presence of the target
%where $\hat a_S^\mathrm{in}$ is the signal mode sent to target region,
%where $G_{S}^\mathrm{det}$ is the gain of the amplifier used to amplify the reflected signal from target.
%, and $\hat a_{n,1}$ is the annihilation operator of the background-noise mode at the with thermal photon occupancy
%$n_{n,1}=
and for $\eta\ll 1$, the added noise to the signal is $n_{1}=\eta\,G_{S}^{\mathrm{det}}(G_S^{\mathrm{amp}}-1) n_{\mathrm{amp},S}+n_{0}$, whose first term
%$N= G_r\, n_{\mathrm{add,S}}\simeq5 \times 10^8$
is due to the amplifier added noise of the first amplification stage before reaching the target, which exceeds
%, $G_r$ is the total gain of the amplifier chain used from the JPC output to the room-temperature target,
the environmental noise $n_\mathrm{env}$ as well as the signal photon numbers used to probe the target. This implies that in our proof of principle demonstration the optimal classical strategy would actually be based on detecting the presence or absence of the amplifier noise rather than the coherences and correlations of the signal-idler path with the measured $\mathrm{SNR}_\mathrm{passive}= (n_1-n_{0})/(n_{0}+1)\simeq 31.4\,$dB for the chosen gain and receiver noise in our setup. However, for lower noise temperature signal amplifiers and lower gain as well as in longer range applications with increased loss, such a passive signature of the detection scheme will be drastically reduced and eventually disappear in the environmental noise at room temperature.

%= n_{RT}+n_{DC}$ where $n_{RT}=672$ is environmental thermal noise of the room-temperature object and $n_{DC}=3.4\times 10^5$ is the noise added after down-conversion.

The final step of the measurement is the application of a digital version of the phase-conjugate receiver~\cite{Guha}. The reflected mode $\hat a_{S,i}^\mathrm{det}$ is first phase-conjugated, and then combined with the 
%calibrated 
idler mode 
%$\hat a_{I}$ 
on a 50-50 beam
splitter. As described in Methods, the SNR of the balanced difference photodetection measurement reads
\begin{equation}\label{SNRq}
\mathrm{SNR}_\mathrm{QI/Cl} =
\frac{(\langle\hat{N}_1\rangle-\langle
\hat{N}_0\rangle)^2}{2\left(\sqrt{(\Delta
N_1)^2} + \sqrt{( \Delta N_0)^2}\right)^2},
\end{equation}
where $\hat N_i=\hat{a}_{i,+}^\dagger\hat{a}_{i,+}-\hat{a}_{i,-}^\dagger\hat{a}_{i,-}$ with $\hat{a}_{i,\pm} =( \hat a_{S,i}^{\mathrm{det}\dagger}+\sqrt{2} \hat a_v \pm \hat{a}_I^{\mathrm{det}})/\sqrt{2}$ is the annihilation operator of the mixed signal and idler modes at the output of the beam splitter in the absence ($i=0$) and the presence ($i=1$) of the target (here $\hat a_v$ is the vacuum noise operator). For the raw SNR without idler calibration we use Eq.~(\ref{SNRq}). In order to simulate perfect photon number detection of the idler mode directly at the JPC output we reduce the variance in the denominator of Eq.~(\ref{SNRq}) by the calibrated idler vacuum and amplifier noise as $\langle \hat a_I^\dagger \hat a_I\rangle = \langle \hat a_i^{\mathrm{det} \dagger} \hat a_i^{\mathrm{det}}\rangle / G_I - (n_{\mathrm{add},I}+1)$, see Methods.

%$\hat{a}_{i,\pm} =( \hat a_{S,i}^{\mathrm{det}\dagger}+\sqrt{2} \hat a_v \pm \hat{a}_I^{\mathrm{det}})/\sqrt{2}$, i.e.~the denominator in Eq.~(\ref{SNRq}) in this case includes the full variance of both detected signals including all vacuum and amplifier noises.}

% Here $\langle \hat O_i\rangle$ and $\langle \Delta \hat{O}_i ^2\rangle$ are the means and variances of the operator $\hat{O}_i$, respectively.

The experiment of coherent
state illumination is performed by generating a weak coherent tone
using a microwave source at room temperature followed by low temperature chain of thermalized attenuators inside the dilution refrigerator. The center frequency
of the coherent tone is $\omega_S$, exactly matched with
the frequency of the signal used in the QI and CI experiments. The coherent tone is
reflected back from the unpumped JPC and directed into the same
measurement chain identical to that of QI and CI, see Fig. \ref{Fig1}(b). The signal is sent to probe a target region and the
detected radiation $\hat a_{S,i}^{\mathrm{det}}$ is used to calculate the SNR of the digital homodyne and heterodyne detections for the same probe power, bandwidth and amplifier noise.

In the absence of a passive signature due to signal noise amplification, digital homodyne detection of a coherent state represents the optimal classical strategy in terms of the SNR, which is given by
\begin{equation}
\mathrm{SNR}_{\mathrm{CS}}^{\mathrm{hom}}=\frac{(\langle \hat X_{S,1}^\mathrm{det}\rangle-\langle \hat X_{S,0}^\mathrm{det}\rangle)^2}{2\Big(\sqrt{(\Delta X_{S,1}^{\mathrm{det}})^2}+\sqrt{(\Delta X_{S,0}^{\mathrm{det}})^2} \Big)^2},
\end{equation}
while the SNR of the digital heterodyne detection is lower and given by
\begin{equation}\label{SNRc}
\mathrm{SNR}_{\mathrm{CS}}^{\mathrm{het}}=\frac{(\langle \hat X_{S,1}^\mathrm{det}\rangle-\langle \hat X_{S,0}^\mathrm{det}\rangle)^2+(\langle \hat P_{S,1}^\mathrm{det}\rangle-\langle \hat P_{S,0}^\mathrm{det}\rangle)^2}{2\Big(\sum\limits_{i=1}^{2}\sqrt{(\Delta X_{S,i}^{\mathrm{det}})^2+(\Delta P_{S,i}^{\mathrm{det}})^2} \Big)^2},
\end{equation}
where $\hat X_{S,i}^{\mathrm{det}}=\frac{\hat a_{S,i}^{\mathrm{det}}+\hat a_{S,i}^{\mathrm{det}\dagger}}{\sqrt{2}}$ and $\hat P_{S,i}^{\mathrm{det}}=\frac{\hat a_{S,i}^{\mathrm{det}}-\hat a_{S,i}^{\mathrm{det}\dagger}}{i\sqrt{2}}$ are the field quadrature operators (see Methods for more details). 
%For many pulses, the cost of the tomographic reconstruction is negligible and these digital detectors perform as the corresponding physical detectors directly applied to the entire ensemble of output states. In particular, the digital homodyne receiver allows one to implement an equivalent physical homodyne detector over the correct rotated quadrature, even when this information about the phase is not known.

%Note that the coherent state \textcolor{purple}{with homodyne} measurement represents the optimal strategy in classical radar systems \cite{Skolnik}.

In Fig.~2(b) we compare the SNR of QI and CI with and without idler calibration for a perfectly reflective object in a zero loss channel $\eta=1$. For comparison, we also include the results of coherent-state illumination with homodyne and heterodyne detection. In all cases the signal mode at room temperature is overwhelmed with amplifier noise. %In Fig.~2 
We use 3 sets of measurements to calculate the standard deviation of the mean SNR of a single mode measurement with measurement time $T=1/B=5\,\mu$s. Each set is based on $M=380\,$k samples ($192\,$k for the coherent state detection) corresponding to a measurement time of 1.87 seconds (0.93 seconds for the coherent state detection). To get the total statistics the measurement time takes 5.6 seconds (2.8 seconds for the coherent state detection). 
For the same measurement bandwidth and using the raw data of the measured quadrature pairs (solid lines) QI (blue dots) outperforms sub-optimum symmetric CI (red dots) by up to 3 dB at low signal photon numbers but it cannot compete with the SNR obtained with coherent state illumination (yellow and green dots). Under the assumption of perfect idler photon number detection, i.e. applying the calibration discussed above (dashed lines), the SNR of QI is up to 4 dB larger than that of symmetric CI and coherent-state illumination with heterodyne detection, which does not require phase information, over the region where the outputs of the JPC are entangled. For signal photon numbers $N_S > 4.5$, where there is no entanglement present in the signal source, the sensitivity of the coherent-state transmitter with heterodyne detection outperforms QI and CI, confirming the critical role of entanglement to improve the sensitivity of the detection. 
%Remarkably, QI is also able to outperform coherent-state illumination with homodyne detection

\begin{figure*}[t]
\centering
\includegraphics[width=1.7\columnwidth]{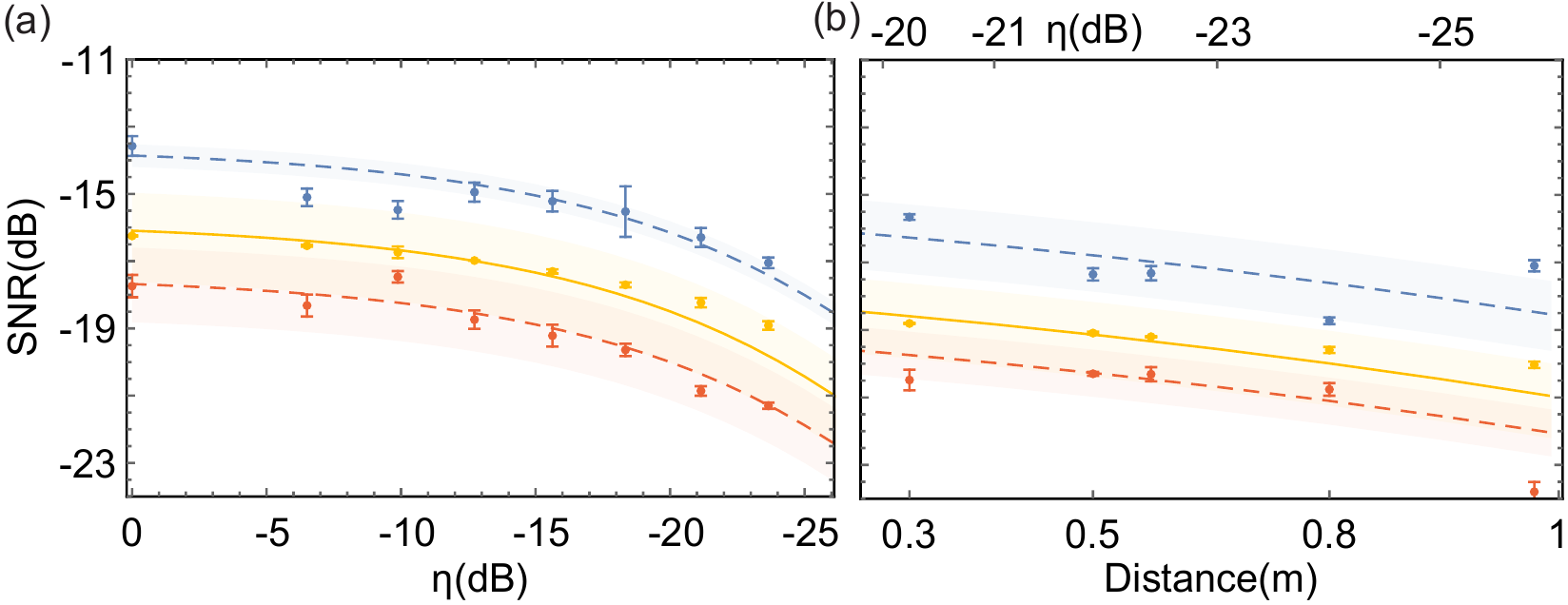}
\caption{\textbf{Low reflectivity quantum correlated noise radar.} The inferred signal to noise ratio (SNR) of calibrated QI (blue) and symmetric CI (red), and the measured coherent-state illumination with digital heterodyne detection (yellow) as a function of (a) the total signal loss $\eta$ and (b) object distance from the transmitting and receiving antennas for free space illumination. The error bars are calculated similar to Fig.~2. For both panels the signal photon number is $N_S=0.5$. The shaded regions are the theoretical uncertainties extracted by fitting the experimental data. The SNR of the coherent state with homodyne detection is not presented in this figure since the expected advantage at the chosen $N_S$ is smaller than systematic errors in this measurement. 
%the as at low photon number and large signal loss the data are too noisy.
%, and the optimal homodyne phase is not easy to find even with larger statistics.  
} \label{circulator}
\end{figure*}

QI with a phase conjugate receiver is potentially able to outperform coherent-state illumination with homodyne detection by up to 3 dB, i.e. the optimum classical benchmark, in the regime of low signal photon numbers. In the region  $N_S<0.4$ the experimentally inferred SNR of QI is approximately 1~dB larger, in agreement with the theoretical prediction taking into account experimental non-idealities like the finite squeezing of the source. In practice though, i.e.~without the applied idler calibration, the quantum advantage compared to coherent homodyne detection
%requires direct photodetection and 
is not accessible with a digital receiver based on heterodyne measurements, even in the case of quantum limited amplifiers, due to the captured idler vacuum noise, which lowers the optimal SNR by at least 3 dB \cite{Tan08,Weedbrook2016}. The experimental results (dots) are in very good agreement with the theoretical prediction (solid and dashed lines). For the theory we rewrite the SNRs Eqs.~(\ref{SNRq})-(\ref{SNRc}) in terms of the signal photon number $N_S=\langle \hat a_S^\dagger \hat a_S\rangle$, the idler photon number $N_I=\langle \hat a_I^\dagger \hat a_I\rangle$, and the signal-idler correlation $\langle \hat a_S \hat a_I\rangle$ at the output of the JPC. These parameters are extracted from the measured and calibrated data as a function of the JPC pump power. %, consistent with the entanglement parameter $\Delta_\mathrm{EPR}$.
Together with the known system gain and noise we plot the theoretical predictions of the various protocols at room-temperature.

An important feature of a radar or short range scanner is its resilience with respect to signal loss. To verify this, as shown in Fig.~1(b), we use two microwave switches at room temperature in the signal line in order to select between a digitally controllable step attenuator to mimic an object with tunable reflectivity and a proof of principle radar setup. With this setup we determine the effects of loss and object reflectivity as well as target distance on the efficiency of the quantum enhanced radar. In Fig.~3(a) we plot the measured SNR of QI, CI and coherent-state illumination with heterodyne detection, as a function of the imposed loss on the signal mode. The calibrated QI protocol is always superior to calibrated symmetric CI and coherent-state illumination with heterodyne detection for a range of effective loss -25~dB $ < \eta < 0$~dB. The dashed lines are the theory predictions from Eq.~(\ref{SNRq}) and Eq.~(\ref{SNRc}) for a fixed chosen signal photon number $N_S=0.5$. The shaded regions represent the confidence interval extracted from the standard deviation of the measured idler photon numbers and the cross-correlations as a function of $\eta$.

In the context of radar, small improvements in the SNR lead to the
%orders of magnitude 
exponentially improved error probability $\mathcal{E}=1/2\, \mathrm{erfc}{(\sqrt{\textrm{SNR} \cdot M})}$, where $M = T_{\mathrm{tot}} B$ is the number of single mode measurements, and $T_\mathrm{tot}$ is the total measurement time required for a successful target detection.
%, in particular for small SNRs that require many copies $M$ for a successful target detection.} 
To test the principle of microwave QI in free-space at room temperature, we amplify and send the microwave signal emitted from JPC to a horn antenna
%with 20~dB antenna gain
and a copper plate representing the target at a variable distance.
%a horn-shape transmitting antenna with 20 dB gain. In the room temperature, we place a movable copper sheet plate as a reflective object in different distances away from the antenna.
The reflected signal from this object is collected using a second antenna of the same type, down-converted, digitized, and combined with the calibrated idler mode to calculate the SNR of the binary decision. With this setup we repeat the measurement for CI and coherent state illumination with heterodyne. Fig.~3(b) shows the SNR of these protocols as a function of the object distance from the transmitting antenna as well as the total loss of the free space link. Calibrated QI reveals higher sensitivity for a reflective target up to 1 meter away from the transmitting antenna. The results are in good agreement with the theoretical model.

\section*{Conclusion}
In this work we have studied proof of concept quantum illumination in the microwave domain, the most natural frequency range for target detection. Assuming perfect idler photon number detection we showed that a quantum advantage is possible despite the entanglement-breaking signal path. Since the best results are achieved for less than one mean photon per mode, our experiment indicates the potential of QI as non-invasive scanning method, e.g. for biomedical applications, imaging of human tissues or non-destructive rotational spectroscopy of proteins, besides its potential use as short-range low-power radar, e.g. for security applications.
%, for which we have achieved a quantum advantage of up to 1~dB over coherent state with homodyne detection and 4 dB over classical correlated and coherent state with heterodyne detection strategies at low signal photon numbers $< 0.4$ obtained with a measurement bandwidth of 200~kHz.
% in closed and populated environments.
%However, in the current demonstration the amplifier noise exceeds the optimal amount of signal photons used to detect the object by about a factor of $\sim$20, which means that 
However, for this initial proof of principle demonstration the amplified bright noise in the target region overwhelms the environmental noise by orders of magnitude, which precludes the non-invasive character at short target distances and presents an opportunity to use the presence or absence of the amplifier noise to detect the object with even higher SNR.
%While adding more signal amplification would help to increase the range, it is also true that the mode temperature (currently dominated by amplifier noise) would need to be reduced for most practical applications. 
%In particular, for noninvasive and stealth-type detection, 
The use of 
%high bandwidth 
quantum limited parametric amplifiers \cite{Yurke1989, Beltran2007, Macklin2015} with limited gain, such that the amplified vacuum does not significantly exceed environmental or typical electronic noise at the target, will help to achieve a practical advantage with respect to the lowest noise-figure coherent state heterodyne receivers at room temperature and, up to the vacuum noise, they will also render the idler calibration obsolete. 
%The use of brighter non-classical microwave sources \cite{Grimm19,Rolland19} \textcolor{purple}{without signal amplification} 
The use of sensitive radiometers or microwave single photon detectors \cite{Kono, Besse, Lescanne} at millikelvin temperatures without signal amplification, represents a promising route to achieve an advantage in practical situations and with respect to ideal coherent state homodyne receivers. One advantage of the presented digital implementation of QI is that it does not suffer from the idler storage problem of receivers that rely on analog photodetection schemes, inherently limiting the accessible range when used as a radar. 
%Since the advantage of QI is restricted to small signal photon numbers and low SNRs where measurement statistics is needed (even in single-shot detector realizations), we believe that an efficiently implemented real-time processing digital approach similar to the one presented here also does not pose a major limitation for the future applicability of microwave QI. 
It is an interesting open question what other types of receivers \cite{Zhuang2017} could be implemented in the microwave domain, based on state of the art superconducting circuit technology and digital signal processing.
%Since such analog photodetection schemes rely on storing the idler mode which limits the achievable range the presented digital phase conjugate receiver has its mertis. 	

\noindent\textbf{Acknowledgments}
We thank Chris Wilson and in particular Jeffrey Shapiro for helpful critical comments and discussions. We thank IBM for donating the JPC used in this work, Jerry Chow and Baleegh Abdo for helpful advice, as well as Mariia Labendik and Riya Sett for their contribution to characterizing the JPC properties.
\\
\textbf{Funding:} This work was supported by the Institute of Science and Technology Austria (IST Austria), the European Research Council under grant agreement number 758053 (ERC StG QUNNECT) and the EU's Horizon 2020 research and innovation programme under grant agreement number 862644 (FET Open QUARTET). S.B. acknowledges support from the Marie Sk\l{}odowska Curie fellowship number 707438 (MSC-IF SUPEREOM), DV acknowledge support from EU's Horizon 2020 research and innovation programme under grant agreement number 732894 (FET Proactive HOT) and the Project QuaSeRT funded by the QuantERA ERANET Cofund in Quantum Technologies, and J.M.F from the Austrian Science Fund (FWF) through BeyondC (F71), a NOMIS foundation research grant, and the EU's Horizon 2020 research and innovation programme under grant agreement number 732894 (FET Proactive HOT).
\\
\noindent\textbf{Author Contributions:}
S.B. S.P., and D.V. proposed and developed the theoretical idea. S.B. and J.M.F conceived the experiment, built the experimental setup and analyzed the data. S.B. performed the theoretical calculation and the measurements. All authors contributed to the manuscript.
\\
\noindent\textbf{Competing interests:} The authors declare no competing interests.

\pagebreak
%\clearpage
\vspace{5cm}
\onecolumngrid

\section*{Supplementary Information}

\section{Josephson parametric converter}
%%%%%%%%%%%%%%%%%%%%%%%%%%%%%%%%%%%%%%%%%%%%%%%SI figures

\begin{figure}[ht]
\centering
\includegraphics[width=3.5in]{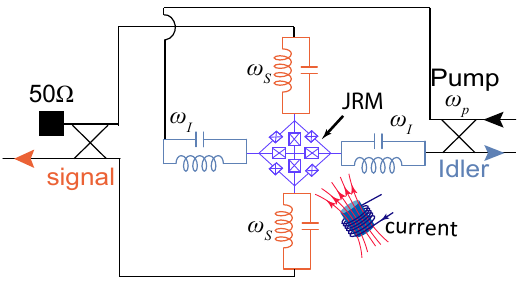}
\caption{\textbf{Schematic representation of the Josephson parametric amplifier (JPC)}. The JPC contains a Josephson ring modulator (JRM) consisting of four Josephson junctions, and four large Josephson junctions inside
the ring act as a shunt inductance for the JRM \cite{Abdo17}. Two microwave resonators are coupled to the JRM forming idler and signal resonators with resonance frequencies $\omega_I$ and $\omega_S$, respectively. These resonators are capacitively coupled to the input and output ports. In order to use the JPC in the three-wave mixing condition the device is biased using an external magnetic field and pumped at frequency  $\omega_p=\omega_I+\omega_S$. Two broadband 180 degree hybrids are used to feed-in and-out the pump, idler, and signal. In this configuration the second port of the signal is terminated using a $50\,\Omega$ cold termination. }
\label{JPC}
\end{figure}

We use a nondegenerate three-wave mixing Josephson parametric converter (JPC) that acts as a nonlinear quantum-limited amplifier whose signal, idler and pump ports are spatially separated, as shown in Fig.~\ref{JPC}. The nonlinearity of the JPC originates from a Josephson ring modulator
(JRM) consisting of four Josephson junctions arranged on a rectangular ring and four large shunting Josephson junctions inside
the ring \cite{Abdo17}. The total geometry supports two differential and one common mode. The correct bias point is selected by inducing a flux in the JRM loop by using an external magnetic field. The two pairs of the microwave half-wavelength microstrip transmission line resonators connected to the center of JRM serve as signal and idler microwave resonators. These resonators are coupled to two differential modes of the JRM and capacitively attached to two external feedlines, coupling in and out the microwave signal to the JPC.

The entanglement between signal mode with frequency $\omega_S$ and idler mode with frequency $\omega_I$ is generated by driving the non-resonant common mode of the JRM at frequency  $\omega_p=\omega_I+\omega_S$. Two off-chip, broadband 180 degree hybrids are used to add the idler or signal modes to the pump drive. In our configuration we apply the pump to the idler side and terminate the other port of the signal hybrid with a $50\,\Omega$ cold termination. The frequency of the signal mode is $\omega_S= 10.09$~GHz and the frequency of the idler mode is $\omega_I= 6.8$~GHz. The maximum dynamical bandwidth and gain of our JPC are $20$~MHz and $30$~dB, respectively. The 1~dB compression point corresponds to the power $-128$~dBm at the input of the device at which the device gain drops by 1~dB and the amplifier starts to saturate. The frequency of the signal and idler modes can be varied over 100~MHz span by applying a dc current to the flux line.

\section{Noise calibration}
\begin{figure}[ht]
\centering
\includegraphics[width=5in]{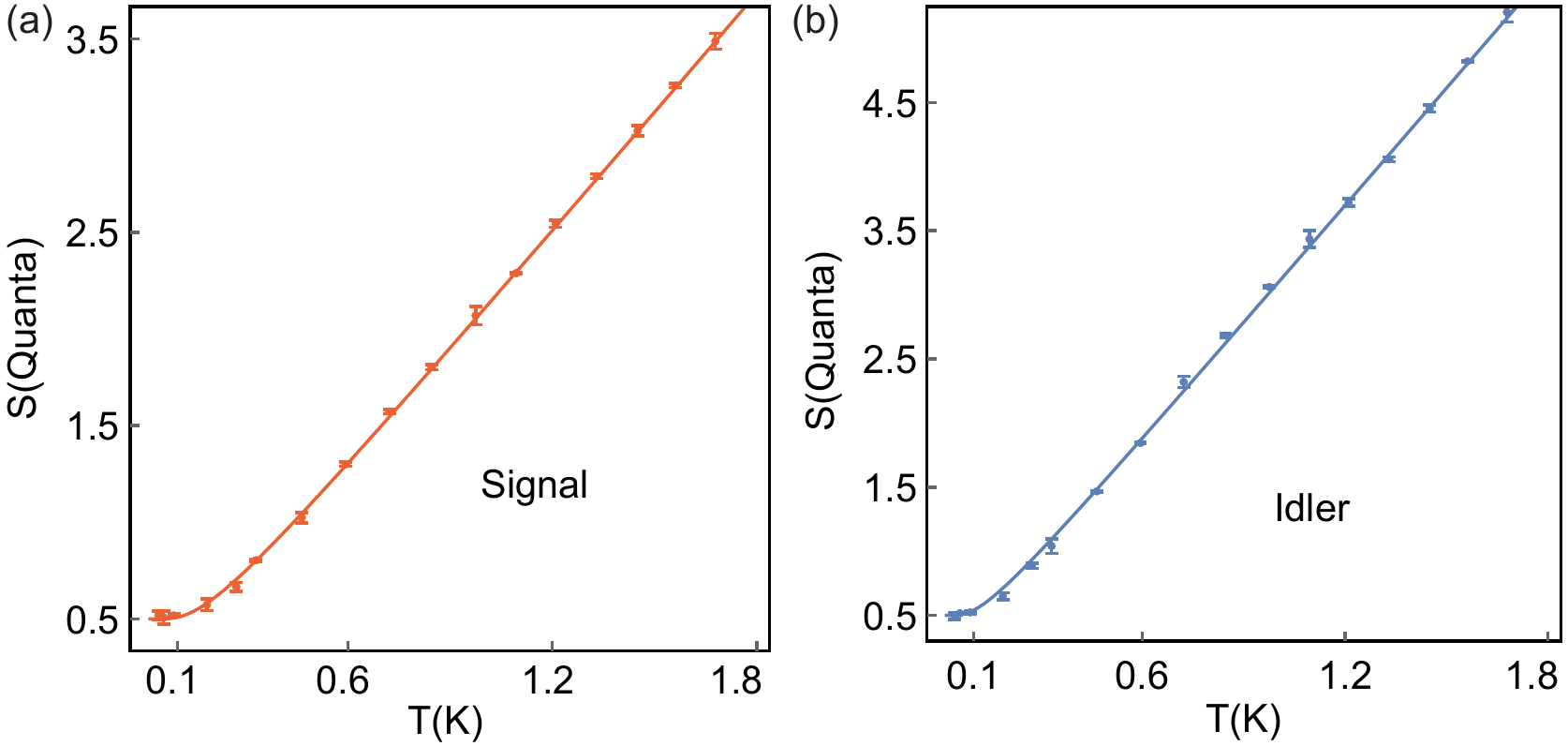}
\caption{\textbf{System noise calibration.} Calibration of signal (a) and idler (b) output channels. The measured noise density in units of quanta, $S_i = N_i/(\hbar \omega_i B R G_i)- n_{\mathrm{add},i}$, is shown as a function of the temperature T of the 50 $\Omega$ load. The error bars indicate the standard deviation obtained from three measurements with $576\,$k quadrature pairs each. The solid lines are fits to Eq.~(5) in units of quanta, which yields the system gain and noise with the standard errors (95$\%$ confidence interval) stated in the main text.}
\label{noisecal}
\end{figure}

The system gain $G_i$ and system noise $n_{\mathrm{add}, i}$ of both signal and idler measurement chains are calibrated
by injecting a known amount of thermal noise using two temperature controlled $50\,\Omega$ cold loads \cite{Flurin2012,Ku2015}. The calibrators are attached to the measurement setup with two copper coaxial cables of the same length and material as the cables used to connect the JPC via two latching microwave switches (Radiall R573423600). A thin copper braid was used for weak thermal anchoring of the calibrators to the mixing chamber plate. By measuring the noise density in V$^2$/Hz at each temperature as shown in Fig.~\ref{noisecal}, and fitting the obtained data with the expected scaling
\begin{equation}\label{noisedensity}
N_i=\hbar \omega_i B R G_i\big[(1/2) \mathrm{coth}[\hbar \omega_i/(2k_BT)]+n_{\mathrm{nadd,i}}\big],
\end{equation}
where $B= 200$ kHz and $R= 50\,\Omega$, we accurately back out the total
gain \beq (G_S,G_I)=(93.98(01),94.25(02))~\mathrm{dB} \eeq  and
the number of added noise photons referenced to the JPC output \beq
(n_{\mathrm{add},S},n_{\mathrm{add},I})=(9.61(04),14.91(1)).\eeq
The 95\% confidence values are taken from the
standard error of the fit shown in Fig.~\ref{noisecal}.

\section{Measurement chain: gain and added noise}

\begin{figure}[ht]
\centering
\includegraphics[width=5in]{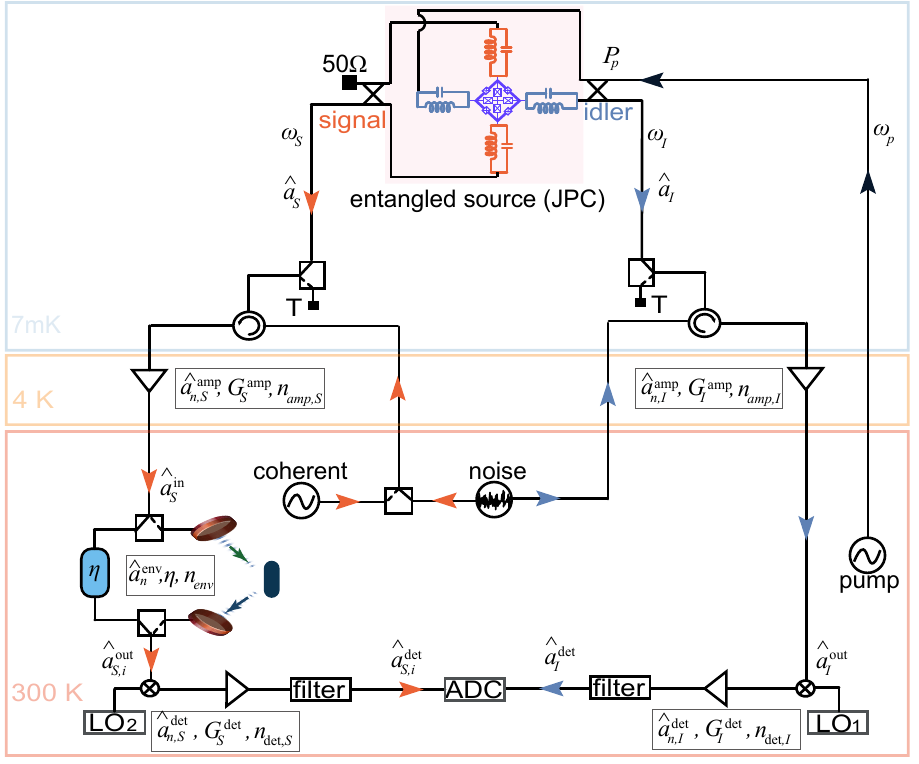}
\caption{\textbf{Full measurement setup}. The outputs of the JPC are amplified in different stages before being down-converted to 20 MHz using two local oscillators ($\mathrm{LO}_1$ and $\mathrm{LO}_2$). After the down-conversion the signals are filtered and amplified once more and then digitized using an analog to digital converter (ADC). Classically-correlated illumination (CI) is performed by using correlated white noise generated by an arbitrary wave form (noise) generator. For coherent-state illumination, we generate a coherent tone and send it to the refrigerator. The signal is reflected from the unpumped JPC and passes through the measurement chain.  %\textcolor{purple}{To perform the heterodyne measurements we first down converting the signal and idler individually and then they are amplified using room-temperature amplifiers with gain $G_i^{\mathrm{det}}$ with $i=S,I$.}
}
\label{mesurmentapp}
\end{figure}

In Fig.~\ref{mesurmentapp} we show the full measurement chain used in our experiment. The outputs of the JPC, the signal $\hat a_S$ and the idler $\hat a_I$, pass through two separate superconducting lines and are amplified individually using two high electron mobility transistor (HEMT model LNF) amplifiers at the 4 K temperature stage and amplified once more at room temperature. The total gain of the amplifier chain is $G_{i}^\mathrm{amp}$. The output of the amplifiers for $ G_{i}^\mathrm{amp}\gg 1 $ are
\begin{eqnarray}\label{eq1}
\hat a_{S}^{\mathrm{in}}&=(\sqrt{G_{S}^\mathrm{amp}}\hat a_S+\sqrt{G_{S}^\mathrm{amp}-1}\hat a_{n,S}^{\mathrm{amp}\dagger}),\nonumber\\
\hat a_{I}^{\mathrm{out}}&=(\sqrt{G_{I}^\mathrm{amp}}\hat a_I+\sqrt{G_{I}^\mathrm{amp}-1}\hat a_{n,I}^{\mathrm{amp}\dagger}),
\end{eqnarray}
where $\hat a_{n,i}^{\mathrm{amp}}$ with $i=S,I$ is the annihilation operator of the noise mode added by the HEMT and one additional room temperature amplifier (not shown) and the preceding cable and connector losses. The idler mode is then down-converted to 20 MHz, filtered, amplified using an amplifier with gain $G_I^{\mathrm{det}}$ and noise annihilation operator  $\hat a_{n,I}^{\mathrm{det}}$, and recorded using an 8 bit analog to digital card (ADC). The down-converted and detected idler mode 
%after digitally post-processing FFT 
is related to the idler mode right after the JPC as
\begin{equation}\label{equation3}
\hat a_{I}^{\mathrm{det}}=\sqrt{G_I}\Big(\hat a_I+\sqrt{\frac{G_I^{\mathrm{amp}}-1}{G_I^{\mathrm{amp}}}}\hat a_{n,I}^{\mathrm{amp\dagger}}+\sqrt{\frac{G_I^{\mathrm{det}}-1}{G_I}}\hat a_{n,I}^{\mathrm{det\dagger}}\Big),
\end{equation}
where $G_{I}=G_I^{\mathrm{det}}G_I^{\mathrm{amp}}=94.25(02)$~dB  is the total gain and
\begin{equation}\label{equation3}
n_{\mathrm{add},I}=\frac{G_I^{\mathrm{amp}}-1}{G_I^{\mathrm{amp}}}\Big(\langle\hat a_{n,I}^{\mathrm{amp\dagger}} \hat a_{n,I}^{\mathrm{amp}}\rangle+1\Big)+\frac{G_I^{\mathrm{det}}-1}{G_I^{\mathrm{amp}}G_I^{\mathrm{det}}}\Big(\langle\hat a_{n,I}^{\mathrm{det\dagger}}\hat a_{n,I}^{\mathrm{det}}\rangle+1\Big)=14.91(1),
\end{equation}
%is the annihilation operator of the total noise added \textcolor{purple}{along} the entire measurement chain with $n_{\mathrm{add},I}=\langle\hat a_{\mathrm{n},I}^\dagger \hat a_{\mathrm{n},I}\rangle=14.91(1)$ 
are the total added noise quanta referenced to the JPC output.

The signal mode is used to probe the target region. The reflected signal from the target region in the presence $H_1$ or absence $H_0$ of the target, respectively, is given by
\begin{subequations}
\begin{eqnarray}\label{eq3}
\hat a_{S,1}^{\mathrm{out}}&=&\sqrt{\eta}\hat a_{S}^{\mathrm{in}}+\sqrt{1-\eta}\,\hat a_{n}^{\mathrm{env}}\,  \mathrm{(hypothesis}\,\, H_1)\\
\hat a_{S,0}^{\mathrm{out}}&=&\hat a_{n}^{\mathrm{env}}\qquad  \qquad \qquad \quad  \mathrm{~(hypothesis}\, H_0),\label{eq3bb}
\end{eqnarray}
\end{subequations}
where $\eta$ is the total signal loss and $\hat a_n^{\mathrm{env}}$ is the annihilation operator of the environmental noise mode at room temperature. In the case of free space illumination we realize the absence of the target by removing the target in front of the antennas, while in the case of using a step-attenuator we mimic the absence of the target by using a 50 $\Omega$ load at the RF port of the mixer. 

The signal mode after down conversion is given by
\begin{equation}\label{eq4}
\hat a_{S,i}^{\mathrm{det}}=(\sqrt{G_{S}^{\mathrm{det}}}\hat a_{S,i}^{\mathrm{out}}+\sqrt{G_{S}^{\mathrm{det}}-1}\hat a_{n,S}^{\mathrm{det}\dagger}),
\end{equation}
with $i=0,1$, $G_{S}^{\mathrm{det}}$ is the gain and $\hat a_{n,S}^{\mathrm{det}}$ is the noise operator of the amplification stage after down conversion. Substituting Eqs. (\ref{eq1}),  (\ref{eq3}) and~(\ref{eq3bb})  into Eq. (\ref{eq4}) gives the detected signal mode in the target-presence
\begin{equation}\label{eq5}
\hat a_{S,1}^{\mathrm{det}}=\sqrt{G_S}\Big(\sqrt{\eta}\hat a_S+\sqrt{\frac{\eta(G_S^{\mathrm{amp}}-1)}{G_S^{\mathrm{amp}}}}\hat a_{n,S}^{\mathrm{amp\dagger}}+\sqrt{\frac{1-\eta}{G_S^{\mathrm{amp}}}}\hat a_{n}^{\mathrm{env}}+\sqrt{\frac{G_S^{\mathrm{det}}-1}{G_S}}\hat a_{n,S}^{\mathrm{det\dagger}}\Big),
\end{equation}
or target-absence
\begin{equation}\label{eq6}
\hat a_{S,0}^{\mathrm{det}}=\sqrt{G_{S}^{\mathrm{det}}}\Big(\hat a_{n}^{\mathrm{env}}+\sqrt{ 1-\frac{1}{G_S^{\mathrm{det}}}}\hat a_{n,S}^{\mathrm{det}\dagger}\Big),
\end{equation}
%where
%\begin{equation}
%\textcolor{blue}{\hat a_{n,1}=\frac{\sqrt{\eta G_{S}^{\mathrm{det}}( G_S^{\mathrm{amp}}-1)}\hat a_{S}^{\mathrm{amp}}+\sqrt{ G_{S}^{\mathrm{det}}(1-\eta)}\,%\hat a_{n}^{\mathrm{env}\dagger}+\sqrt{ G_S^{\mathrm{det}}-1}\hat a_{n,S}^{\mathrm{det}}}{\sqrt{\eta G_{S}-1}},}
%\end{equation}
%is the annihilation operators of the total noise added in presence of the target. 
where $G_{S}=G_S^{\mathrm{det}}G_S^{\mathrm{amp}}=93.98(01)$~dB is the total gain with $G_S^{\mathrm{det}}=16.82$~dB, $G_S^{\mathrm{amp}}=77.16$~dB, and
\begin{equation}\label{equation3}
n_{\mathrm{add},S}=\frac{G_S^{\mathrm{amp}}-1}{G_S^{\mathrm{amp}}}\Big(\langle\hat a_{n,S}^{\mathrm{amp\dagger}} \hat a_{n,S}^{\mathrm{amp}}\rangle+1\Big)+\frac{G_S^{\mathrm{det}}-1}{G_S^{\mathrm{amp}}G_S^{\mathrm{det}}}\Big(\langle\hat a_{n,S}^{\mathrm{det\dagger}}\hat a_{n,S}^{\mathrm{det}}\rangle+1\Big)=9.61(04),
\end{equation}
are the total added noise quanta at the JPC output. The total added noise in the presence of the target is given by 
$n_{1}=\eta\,G_S^{\mathrm{det}}(G_S^{\mathrm{amp}}-1) n_{\mathrm{amp},S}+(1-\eta)G_S^{\mathrm{det}}n_{\mathrm{env}}+(G_S^{\mathrm{det}} - 1)n_{\mathrm{det},S}$ which, in the limit of $\eta \ll 1 $, leads to $n_{1}=\eta\,G_S^{\mathrm{det}}(G_S^{\mathrm{amp}}-1) n_{\mathrm{amp},S}+n_{0}$, where $(G_S^{\mathrm{amp}}-1) n_{\mathrm{amp},S}\approx 5\times 10^8$. The total added noise in the absence of the target is $n_{0}= G_S^{\mathrm{det}} n_\mathrm{env}+(G_S^{\mathrm{det}}- 1)n_{\mathrm{det},S}$, where $n_\mathrm{env}=\langle \hat a_{n}^{\mathrm{env}\dagger}\hat a_{n}^{\mathrm{env}}\rangle= 672$ is the environmental thermal noise of the room-temperature object and $n_{\mathrm{det},S}=\langle \hat a_{\mathrm{n},S}^{\mathrm{det}\dagger}\hat a_{\mathrm{n},S}^{\mathrm{det}}\rangle+1\approx 3\times 10^5$ is the receiver noise dominated by the amplifier noise after down-conversion to the intermediate frequency.

\section{Digital post-processing}

In this section we explain how the single-mode post-processing was performed. As shown in Fig.~\ref{posrprocess}(a), the down-converted and amplified signal and idler modes are continuously recorded with 100 MS/s using a two-channel ADC with 8 bit resolution. The total measurement time of the QI/CI detections (coherent-state detections) is $5.76$ seconds ($2.88$ seconds) in which the recorded data are chopped to $M=1.15\times10^6\,(6 \times 10^5)$ records, each contains $500$ samples which corresponds to a filter bandwidth of 200 kHz. The $500$ samples are used to perform fast Fourier transform (FFT) on each record individually and extract the complex quadrature voltages $I_I$, $Q_I$ and $I_S$, $Q_S$ of the intermediate frequency component at 20~MHz.
We calculate the detected field quadratures of both signal and idler modes $X_i^{\mathrm{det}}=I_i/\sqrt{\hbar\omega_i B R} $ and $ P^{\mathrm{det}}_i =Q_i/\sqrt{\hbar\omega_i B R}$ with $i=S,I$ for $M$ measurement results, which have the same measurement statistics as the quadrature operators $\hat X_i^{\mathrm{det}}$ and $\hat P_i^{\mathrm{det}}$, where 
%These quadratures are then used to calculate the covariances of the signal and idler modes, 
$\hat a^{\mathrm{det}}_i=(\hat X^{\mathrm{det}}_i+\mathrm{i}\, \hat P^{\mathrm{det}}_i)/\sqrt{2}$. 
%\textcolor{purple}{In the case of the simulated ideal idler photon number detection}, we apply the calibration $\langle \hat a_I^\dagger \hat a_I\rangle = \langle \hat a_i^{\mathrm{det} \dagger} \hat a_i^{\mathrm{det}}\rangle / G_I - (n_\mathrm{add,I}+0.5)$.
%, \textcolor{purple}{which in practice is obtained from successive measurements with the signal on and off.} The results of this post-processing step are used to calculate the SNRs.  
%\textcolor{purple}{In the case of the shown raw SNRs of CI and QI, the full detected variance is taken into account and no idler calibration is applied.}

\section{Digital phase-conjugate receiver: QI and CI}

\begin{figure}[ht]
\centering
\includegraphics[width=6.5in]{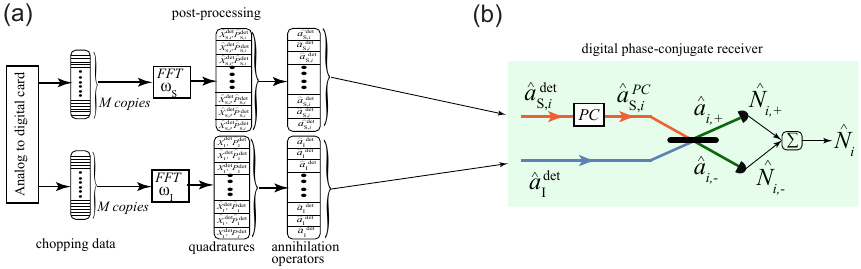}\caption{\textbf{Schematic of the post-processing.} (a) The recorded data from the ADC is chopped in $M$ shorter arrays. We apply digital FFT at idler ($\omega_I$) and signal frequencies ($\omega_S$) after analog downconversion on each array individually to infer the measurement statistics of the signal and idler mode quadratures $\hat X_i^\mathrm{det} $ and $ \hat P^\mathrm{det}_i $  with $i=S,I$. The measurement results are then used to calculate the covariances of the signal and idler modes $\hat a^\mathrm{det}_i=(\hat X^\mathrm{det}_i+\mathrm{i}\, \hat P^\mathrm{det}_i)/\sqrt{2}$. %To \textcolor{purple}{infer the SNR we use the calibrated} idler mode using the system gain and noise. 
(b) The digital phase-conjugate receiver used to infer the SNR of QI and CI. The $M$ copies of the signal and idler modes, generated in post-processing, are sent one by one to the digital phase-conjugate receiver.  A 50-50 beam splitter mixes the phase conjugated signal mode $\hat a_{S,i}^\mathrm{PC}$ returned from target region, with the locally detected idler mode $\hat a_I^\mathrm{det}$. The beam splitter's outputs are detected, yielding classical outcomes equivalent to the quantum measurements $\sum_{k=1}^M\hat{N}_{i,\pm}^{(k)}$ (includes all $M$ copies), and the difference of these outputs, equivalent to the quantum measurement of $\hat{N}_i$, is used as the input to a threshold detector (not shown) whose output is the target absence or presence decision.}
\label{posrprocess}
\end{figure}

Both the JPC and a correlated classical source generate a zero-mean, two-mode Gaussian state
with a nonzero 
%phase-sensitive 
cross correlation $\langle \hat a_S\hat a_I\rangle=\langle \hat a_S^\mathrm{det}\hat a_I^\mathrm{det}\rangle/\sqrt{G_S G_I}$. To quantify this correlation, $M$ copies of the measurement results with the statistics of $\hat a_S^\mathrm{det}$ and $\hat a_I^\mathrm{det}$
% , generated in post-processing, 
are sent individually through the digital phase-conjugate receiver in which we first perform phase-conjugation on the received individual signal $\hat a_{S,i}^{PC}=\sqrt{2}\hat a_v+\hat{a}_{S,i}^{\mathrm{det}\dagger}$ ($\hat a_v$ is the vacuum operator) and then mix it with the retained corresponding idler modes on a 50--50 beam splitter, as shown in Fig.~\ref{posrprocess}(b), whose outputs are
\begin{equation}
\hat{a}_{i,\pm} \equiv \frac{\hat{a}_{S,i}^{PC} \pm \hat{a}_I^\mathrm{det}}{\sqrt{2}}.
\end{equation}

%Note that \textcolor{purple}{for the raw SNR without idler calibration $\hat{a}_{i,\pm} =( \hat{a}_{S,i}^{PC}\pm \hat{a}_I^{\mathrm{det}})/\sqrt{2}$.} 
The target absence-or-presence decision is made by comparing the difference of the two detectors' total photon counts  \cite{Barzanjeh15}, which is equivalent to the measurement of the operator
\begin{equation}
\hat{N}_i = \hat{N}_{i,+}-\hat{N}_{i,-}.
\end{equation}
where $\hat{N}_{i,\pm} \equiv \hat{a}_{i,\pm}^{\dagger}\hat{a}_{i,\pm}$. Since our QI protocol employs a large number of copies $M$, the central limit theorem implies that the measurement of $\sum_{k=1}^M\hat{N}_{i,\pm}^{(k)}$ yields a random variable that is Gaussian, conditioned on target absence or target presence. It follows that the 
%\textcolor{purple}{inferred} 
receiver's SNR for QI or CI satisfies
\begin{equation}\label{rawsnr}
\mathrm{SNR}_\mathrm{QI/Cl} =
\frac{(\langle\hat{N}_1\rangle-\langle
\hat{N}_0\rangle)^2}{2\left(\sqrt{(\Delta
N_1)^2} + \sqrt{(\Delta N_0)^2}\right)^2},
\end{equation}
with $\langle\hat{O}_i\rangle$ and $(\Delta O_i)^2=\langle\hat{O}_i^2\rangle-\langle\hat{O}_i\rangle^2$, for $i = 0,1$, being the conditional means and conditional variances of $\hat{O}_i$, respectively, and the brackets $\langle...\rangle$ denote an average over all of the $M$ copies. For the reported raw SNR we use Eq.~(\ref{rawsnr}) without any calibration applied. To infer the hypothetical SNR that could be obtained with access to the idler mode directly at the JPC output $\hat{a}_{I}$, we rewrite Eq.~(\ref{rawsnr}) in terms of single-mode moments, i.e.,
\begin{equation}
\label{SNR} \mathrm{SNR}_\mathrm{QI/Cl} =
\frac{[(\langle\hat{N}_{1,+}\rangle-\langle
\hat{N}_{1,-}\rangle)-(\langle\hat{N}_{0,+}\rangle-\langle
\hat{N}_{0,-}\rangle)]^2}{2\left(\sqrt{(\Delta
N_1)^2} + \sqrt{(\Delta N_0)^2}\right)^2},
\end{equation}
where
\begin{subequations}
\begin{eqnarray}
\langle \hat{N}_{0,+}\rangle - \langle \hat{N}_{0,-}\rangle &=& 0,\\
\langle \hat{N}_{1,+}\rangle - \langle \hat{N}_{1,-}\rangle &=&
2\sqrt{\eta\,G_S}\langle \hat a_S \hat a_I\rangle,
\end{eqnarray}
\end{subequations}
and \cite{Guha}
\begin{equation}\label{inferredSNR}
(\Delta N_i)^2  =
\langle \hat{N}_{i,+}\rangle(\langle \hat{N}_{i,+}\rangle +1) +
\langle \hat{N}_{i,-}\rangle(\langle \hat{N}_{i,-}\rangle+1)
 - (\langle \hat{a}_{S,i}^{PC\dagger}\hat{a}_{S,i}^{PC}\rangle - \langle \hat{a}_{I}^{\dagger}\hat{a}_{I}\rangle)^2/2,
\end{equation}
for $i = 0,1$,
%To experimentally determine the idler calibrated SNR we use Eq.~(\ref{SNR}) with the calibrated idler noise variance of Eq.~(\ref{inferredSNR}), 
where we take the calibrated noiseless idler photon number $\langle \hat a_I^\dagger \hat a_I\rangle = \langle \hat a_I^{\mathrm{det} \dagger} \hat a_I^{\mathrm{det}}\rangle / G_I - (n_{\mathrm{add},I}+1)$. Here, $\langle \hat a_S \hat a_I\rangle$ is presumed to be real valued, which in general requires phase information in order to apply the appropriate quadrature rotation that maximizes the signal-idler correlation.

\section{SNR of the coherent state illumination: heterodyne and homodyne measurements}

To perform the coherent state illumination, we generate a coherent signal at room temperature and send it into the dilution refrigerator where the mode $\hat a_{S}$ is reflected at the JPC output port and passes through exactly the same measurement line as in the case of of the QI and CI protocols. The amplified signal mode is then used to probe the target region and measured via heterodyne detection. In the presence of the target the measured signal is given by Eq.~(\ref{eq5}) with $\langle \hat a_{S,1}^\mathrm{det} \rangle=\sqrt{\eta\,G_{S}}\langle \hat a_{S} \rangle $, and in the absence of target it is given by Eq.~(\ref{eq6}) with $\langle \hat a_{S,0}^\mathrm{det}\rangle=0$.
Similar to QI, we perform data-processing on the recorded coherent-state outputs and use $M$ measurement results of the field quadrature operators $\hat X_S^\mathrm{det}$ and $\hat P_S^\mathrm{det}$ to perform a digital heterodyne detection with the following SNR
\begin{equation}
\mathrm{SNR}_{\mathrm{CS}}^{\mathrm{het}}=\frac{(\langle \hat X_{S,1}^\mathrm{det}\rangle-\langle \hat X_{S,0}^\mathrm{det}\rangle)^2+(\langle \hat P_{S,1}^\mathrm{det}\rangle-\langle \hat P_{S,0}^\mathrm{det}\rangle)^2}{2\Big(\sqrt{(\Delta X_{S,1}^{\mathrm{det}})^2+(\Delta P_{S,1}^{\mathrm{det}})^2}+\sqrt{(\Delta X_{S,0}^{\mathrm{det}})^2+(\Delta  P_{S,0}^{\mathrm{det}})^2} \Big)^2}.
\end{equation}
For the digital homodyne detection we use phase information to rotate the signal to the relevant quadrature direction and obtain the improved SNR
\begin{equation}
\mathrm{SNR}_{\mathrm{CS}}^{\mathrm{hom}}=\frac{(\langle \hat X_{S,1}^\mathrm{det}\rangle-\langle \hat X_{S,0}^\mathrm{det}\rangle)^2}{2\Big(\sqrt{(\Delta X_{S,1}^{\mathrm{det}})^2}+\sqrt{(\Delta  X_{S,0}^{\mathrm{det}})^2} \Big)^2}.
\end{equation}


\begin{thebibliography}{30}

\bibitem{SensingReview} S. Pirandola,  B. R. Bardhan, T. Gehring, C. Weedbrook, S. Lloyd, Advances
in Photonic Quantum Sensing. \textit{Nat. Photon.} \textbf{12}, 724-733 (2018).

\bibitem{NielsenChuang}  M. A. Nielsen, I. L. Chuang,  \textit{Quantum computation and quantum information} (Cambridge University Press, Cambridge, 2000).

\bibitem{Weedbrook12} C. Weedbrook, S. Pirandola, R. Garcia-Patron, N. J. Cerf, T. C. Ralph, J. H. Shapiro, S. Lloyd, Gaussian quantum information. \textit{Rev. Mod. Phys.} \textbf{84}, 621-669 (2012).

\bibitem{Hayashi17} M. Hayashi, \textit{Quantum Information Theory: Mathematical Foundation} (Springer-Verlag, 2017).

\bibitem{Watrous18} J. Watrous, \textit{The theory of quantum information} (Cambridge University Press, Cambridge, 2018).

\bibitem{Braunstein94} S. L. Braunstein, C. M. Caves, Statistical distance and the geometry of quantum states. \textit{Phys. Rev. Lett.} \textbf{72}, 3439-3443 (1994).

\bibitem{Schoelkopf08} R. J. Schoelkopf, S. M. Girvin,  Wiring up quantum systems. \textit{Nature} \textbf{451}, 664-669 (2008).

\bibitem{Kimble2008} H. J. Kimble, The quantum internet. \textit{Nature} \textbf{453}, 1023 (2008).

\bibitem{PirBra16} S. Pirandola, S. L. Braunstein, Unite to build a quantum Internet. \textit{Nature} \textbf{532}, 169 (2016).

\bibitem{Wehner18} S. Wehner, D. Elkouss, R. Hanson, Quantum internet: A vision for the road ahead. \textit{Science} \textbf{362}, eaam9288 (2018).

\bibitem{Lloyd08} S. Lloyd, Enhanced sensitivity of photodetection via quantum illumination. \textit{Science} \textbf{321}, 1463-1465 (2008).

\bibitem{Tan08}  Si-Hui Tan, Baris I. Erkmen, Vittorio Giovannetti, Saikat Guha, Seth Lloyd, Lorenzo Maccone, Stefano Pirandola, Jeffrey H. Shapiro, Quantum illumination with Gaussian States. \textit{Phys. Rev. Lett.} \textbf{101}, 253601 (2008).

\bibitem{Lopaeva2013} E. D. Lopaeva, I. Ruo Berchera, I. P. Degiovanni, S. Olivares, G. Brida, and M. Genovese, Experimental Realization of Quantum Illumination. \textit{Phys. Rev. Lett.} \textbf{110}, 153603 (2013).

\bibitem{Zhang2013} Zheshen Zhang, Maria Tengner, Tian Zhong, Franco N. C. Wong, Jeffrey H. Shapiro, Entanglement's Benefit Survives an Entanglement-Breaking Channel. \textit{Phys. Rev. Lett.} \textbf{111}, 010501 (2013).
%

\bibitem{Zhang2015} Zheshen Zhang, Sara Mouradian, Franco N. C. Wong and Jeffrey H. Shapiro, Entanglement-Enhanced Sensing in a Lossy and Noisy Environment. \textit{Phys. Rev. Lett.} \textbf{114}, 110506 (2015).


\bibitem{Weedbrook2016} C. Weedbrook, S. Pirandola, J. Thompson, V. Vedral and M. Gu, How discord underlies the noise resilience of quantum illumination.\textit{New Journal of Physics} \textbf{18}, 043027 (2016).

%
\bibitem{Shapiro2019} Jeffrey H. Shapiro, The Quantum Illumination Story. arXiv:1910.12277 (2019).
 
\bibitem{Choi2016}J. Choi, V. Va, N. Gonzalez-Prelcic, R. Daniels, C. R. Bhat, R. W. Heath, Millimeter-Wave Vehicular Communication to Support Massive Automotive Sensing.  IEEE Communications Magazine, \textbf{54}, 160 (2016).
Page(s): 160 - 167

\bibitem{Lin1992}J.C. Lin, Microwave sensing of physiological movement and volume change: A review. \textit{Bioelectromagnetics} \textbf{13}, 557 (1992).

\bibitem{Wilson19}  C. W. Sandbo Chang,  A. M. Vadiraj, J. Bourassa, B. Balaji, C. M. Wilson, Quantum-enhanced noise radar. \textit{Appl. Phys. Lett.} \textbf{114}, 112601 (2019).

\bibitem{Wilson1903} D. Luong,  C. W. Sandbo Chang, A. M. Vadiraj, A. Damini, C. M. Wilson, B. Balaji, Receiver Operating Characteristics for a Prototype Quantum Two-Mode Squeezing Radar. arXiv:1903.00101.

\bibitem{Guha} Saikat Guha, Baris I. Erkmen, Gaussian-state quantum-illumination receivers for target detection. \textit{Phys. Rev. A} \textbf{80}, 052310 (2009).

\bibitem{Barzanjeh15}  Shabir Barzanjeh, Saikat Guha, Christian Weedbrook, David Vitali, Jeffrey H. Shapiro, and Stefano Pirandola, Microwave quantum illumination. \textit{Phys. Rev. Lett.} \textbf{114}, 080503 (2015).

\bibitem{Bergeal2010} N. Bergeal, R. Vijay, V.E. Manucharyan, I. Siddiqi, R.
Schoelkopf, S. Girvin, and M. Devoret, Phase-preserving amplification near the quantum
limit with a Josephson ring modulator. \textit{Nat. Phys.} \textbf{6}, 296
(2010).

\bibitem{Abdo2013} B. Abdo, A. Kamal, M. H. Devoret, Nondegenerate three-wave mixing with the Josephson ring modulator. \textit{Phys. Rev. B} \textbf{87}, 014508 (2013).

\bibitem{Flurin2012} E. Flurin, N. Roch, F. Mallet, M. Devoret, B. Huard, Generating entangled microwave radiation over two transmission lines. \textit{Phys. Rev. Lett.} \textbf{109}, 183901 (2012).

\bibitem{Silveri2015} Matti Silveri, Evan Zalys-Geller, Michael Hatridge, Zaki Leghtas, Michel H. Devoret, S. M. Girvin, Theory of remote entanglement via quantum-limited phase-preserving amplification. \textit{Phys. Rev. A} \textbf{93}, 062310 (2016).

\bibitem{daSilva10} Marcus P. da Silva, Deniz Bozyigit, Andreas Wallraff, Alexandre Blais, Schemes for the observation of photon correlation functions in circuit QED with linear detectors. \textit{Phys. Rev. A} \textbf{82}, 043804 (2010).

\bibitem{Bozyigit11} D. Bozyigit, C. Lang, L. Steffen, J. M. Fink, C. Eichler, M. Baur, R. Bianchetti, P. J. Leek, S. Filipp, M. P. da Silva, A. Blais and A. Wallraff, Antibunching of microwave-frequency photons observed in correlation measurements using linear detectors. \textit{Nat. Phys.} \textbf{7}, 154 (2011).

%\bibitem{Eichler2011} C. Eichler, D. Bozyigit, C. Lang, L. Steffen, J. Fink, and A. Wallraff, Experimental State Tomography of Itinerant Single Microwave Photons. \textit{Phys. Rev. Lett.} \textbf{106}, 220503 (2011).

\bibitem{Eichler12} C. Eichler, D. Bozyigit, A. Wallraff, Characterizing quantum microwave radiation and its entanglement with superconducting qubits using linear detectors. \textit{Phys. Rev. A} \textbf{86}, 032106 (2012).

\bibitem{Barzanjeh19} S. Barzanjeh, E. S. Redchenko, M. Peruzzo, M. Wulf, D. P. Lewis, G. Arnold, J. M. Fink, Stationary entangled radiation from micromechanical motion. \textit{Nature} \textbf{570}, 480-483 (2019).

\bibitem{Menzel} E. P. Menzel, R. Di Candia, F. Deppe, P. Eder, L. Zhong, M. Ihmig, M. Haeberlein, A. Baust, E. Hoffmann, D. Ballester, K. Inomata, T. Yamamoto, Y. Nakamura, E. Solano, A. Marx, R. Gross, Path entanglement of continuous-variable quantum microwaves. \textit{Phys. Rev. Lett.} \textbf{109}, 250502 (2012).

\bibitem{Duan2000} L.-M. Duan, G. Giedke, J. I. Cirac, P. Zoller, Inseparability criterion for continuous variable systems. \textit{Phys. Rev. Lett.} \textbf{84}, 2722-2725 (2000).
%

\bibitem{Yurke1989} B. Yurke, L. R. Corruccini, P. G. Kaminsky, L. W. Rupp, A. D. Smith, A. H. Silver, R. W. Simon, E. A. Whittaker, Observation of parametric amplification and deamplification in a Josephson parametric amplifier. \textit{Phys. Rev. A} \textbf{39}, 2519-2533 (1989).

%

\bibitem{Beltran2007} M. A. Castellanos-Beltran, K. W. Lehnert, Widely tunable parametric amplifier based on a superconducting quantum interference device array resonator. \textit{Applied Physics Letters} \textbf{91} 083509 (2007)


\bibitem{Macklin2015} C. Macklin, K. O'Brien, D. Hover, M. E. Schwartz, V. Bolkhovsky, X. Zhang, W. D. Oliver, I. Siddiqi, A near-quantum-limited Josephson traveling-wave parametric amplifier. \textit{Science} \textbf{350}, 307 (2015).

%\bibitem{Grimm19} A. Grimm, F. Blanchet, R. Albert, J. Leppakangas, S. Jebari, D. Hazra, F. Gustavo, J.-L. Thomassin, E. Dupont-Ferrier, F. Portier, M. Hofheinz, Bright On-Demand Source of Antibunched Microwave Photons Based on Inelastic Cooper Pair Tunneling. \textit{Phys. Rev. X}, \textbf{9}, 021016 (2019).

%\bibitem{Rolland19} C. Rolland, A. Peugeot, S. Dambach, M. Westig, B. Kubala, Y. Mukharsky, C. Altimiras, H. le Sueur, P. Joyez, D. Vion, P. Roche, D. Esteve, J. Ankerhold, F. Portier, Antibunched Photons Emitted by a dc-Biased Josephson Junction. \textit{Phys. Rev. Lett.}, \textbf{122}, 186804 (2019).

\bibitem{Kono} S. Kono, K. Koshino, Y. Tabuchi, A. Noguchi, Y.
Nakamura, Quantum non-demolition detection of an itinerant microwave photon. \textit{Nature Physics}, \textbf{14}, 546-549, (2018).

\bibitem{Besse} Jean-Claude Besse, Simone Gasparinetti, Michele C. Collodo, Theo Walter, Philipp Kurpiers, Marek Pechal, Christopher Eichler, Andreas Wallraff, Single-shot quantum nondemolition detection of individual itinerant microwave photons, \textit{Phys. Rev. X},\textbf{8}, 021003 (2018).

\bibitem{Lescanne} Raphael Lescanne, Samuel Deleglise, Emanuele Albertinale, Ulysse Reglade, Thibault Capelle, Edouard Ivanov, Thibaut Jacqmin, Zaki Leghtas, Emmanuel Flurin, Detecting itinerant microwave photons with engineered non-linear dissipation. arXiv:1902.05102.

\bibitem{Zhuang2017} Quntao Zhuang, Zheshen Zhang, and Jeffrey H. Shapiro, Optimum Mixed-State Discrimination for Noisy Entanglement-Enhanced Sensing. \textit{Phys. Rev. Lett.},\textbf{118}, 040801 (2017).

%\bibitem {Abdo17} B. Abdo, J. M. Chavez-Garcia, M. Brink, G. Keefe, J. M. Chow, Time-multiplexed amplification in a hybrid-less and coil-less Josephson parametric converter. \textit{Appl. Phys. Lett.} \textbf{110}, 082601 (2017).

%\bibitem{Ku2015} H. S. Ku, W. F. Kindel, F. Mallet, S. Glancy, K. D. Irwin, G. C. Hilton, L. R. Vale, K. W. Lehnert, Generating and verifying entangled itinerant microwave fields with efficient and independent measurements. \textit{Phys. Rev. A} \textbf{91}, 042305 (2015).
\end{thebibliography}

\begin{thebibliography}{30}


\bibitem {Abdo17} B. Abdo, J. M. Chavez-Garcia, M. Brink, G. Keefe, J. M. Chow, Time-multiplexed amplification in a hybrid-less and coil-less Josephson parametric converter. \textit{Appl. Phys. Lett.} \textbf{110}, 082601 (2017).
\bibitem{Flurin2012} E. Flurin, N. Roch, F. Mallet, M. Devoret, B. Huard, Generating entangled microwave radiation over two transmission lines. \textit{Phys. Rev. Lett.} \textbf{109}, 183901 (2012).
\bibitem{Ku2015} H. S. Ku, W. F. Kindel, F. Mallet, S. Glancy, K. D. Irwin, G. C. Hilton, L. R. Vale, K. W. Lehnert, Generating and verifying entangled itinerant microwave fields with efficient and independent measurements. \textit{Phys. Rev. A} \textbf{91}, 042305 (2015).
\bibitem{Guha} Saikat Guha, Baris I. Erkmen, Gaussian-state quantum-illumination receivers for target detection. \textit{Phys. Rev. A} \textbf{80}, 052310 (2009).

\end{thebibliography}
\end{document}